\documentclass{article}
\usepackage{spconf,amsmath,graphicx,hyperref}
\usepackage[caption=false,font=footnotesize]{subfig}

\title{Generative Quantum Machine Learning Methods for Cancer Detection using Augmented Datasets}
\name{\shortstack[c]{
Sophia Knutson$^1$, Nandika Goyal$^2$, Gennaro De Luca$^3$,\\[-2pt]
Glen Uehara$^2$, Tanay Kamlesh Patel$^2$, Andreas Spanias$^2$}
\sthanks{This project is funded by the National Science Foundation (NSF) Award 2349567. Any opinions, findings, conclusions, or recommendations expressed in this material are those of the author(s) and do not necessarily reflect those of the NSF.}}

\address{$^1$ School of Computer Science, Carnegie Mellon University, Pittsburgh, PA, USA\\ 
$^2$ SenSIP Center, Arizona State University, Tempe, AZ, USA \\ 
$^3$ School of Computing and Augmented Intelligence, Arizona State University, Tempe, AZ, USA}
\begin{document}
\ninept
\maketitle
\begin{abstract}
Cancer detection using AI methods and image databases faces several challenges, including limited and imbalanced image datasets, long algorithm training times, and high-complexity feature generation. In particular, many cancer image datasets contain very few malignant examples compared to benign ones, limiting the amount of minority-class data available for training. Although classical AI and especially deep learning methods have had some success in providing cancer diagnostics, Quantum Machine Learning (QML) methods may offer advantages through high-dimensional feature representations, generative data augmentation, and the exploitation of feature correlations through quantum properties such as entanglement. In this paper, we propose using a generative QML method on the compressed breast ultrasound image dataset (BUSI) to generate synthetic images and hence augmented datasets for classifier training. In particular, we train a convolutional neural network with real, synthetic, and a combination of real and synthetic images. We test the algorithm on real images and provide comparative results for all three training cases. Our results show that the classifier trained entirely on synthetic images achieved an AUC-ROC of 0.802 (80.2\%), compared with 0.760 (76.0\%) for the real-data baseline.

\end{abstract}

%
\section{Introduction}
\label{sec:intro}
\vspace{-2mm}
Manual analysis of breast ultrasound images for cancer diagnosis remains both  expensive and time-consuming \cite{venkatesh2015going}. Several image processing (IP) and machine learning (ML) automated methods have been proposed for breast cancer diagnosis. However, there is a persistent need to improve the robustness and accuracy of diagnosis. Several studies used ML and deep learning for breast cancer detection with multiple imaging modalities \cite{mahmood2020brief} including infrared thermal imaging \cite{mambou2018breast}, histopathology and whole-slide images \cite{aboudessouki2025enhanced,lopez2023deep,li2019discriminative,pattichis2002modular}, and mammography \cite{loizidou2025subtraction,liasis2012combination,sun2004two,muralidhar2008computer}. Recent approaches have focused on more challenging aspects of breast cancer analysis, including addressing data imbalance, semi-supervised learning with limited labeled data \cite{jiang2026semi}, attention-based feature fusion \cite{liu2025famf}, and cross-image dependency modeling for segmentation \cite{wu2023cross}. Although promising, the performance of these methods can be limited when training datasets are small or imbalanced \cite{johnson2019survey}.

\begin{figure}[htb]
    \centering
    \includegraphics[width=\columnwidth]{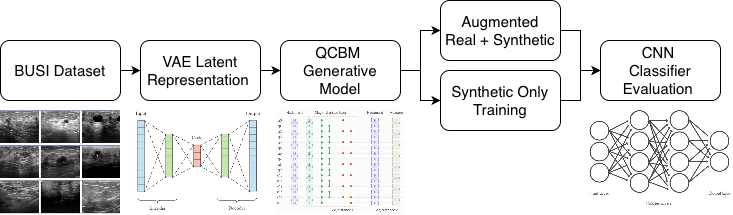}
    \caption{
    Overview of the proposed Quantum Circuit Born Machine (QCBM) method for breast ultrasound data generation, augmentation and classification. The BUSI dataset along with a variational autoencoder (VAE) are used and classification is performed with a convolutional neural network (CNN).
    }
    \label{fig:pipeline_overview}
\end{figure}

Medical imaging datasets, such as the breast ultrasound image dataset (BUSI), are imbalanced, with fewer labelled malignant images than benign or normal ones. Classical approaches address this limitation by generating additional samples for the underrepresented classes, through algorithms like generative adversarial minority oversampling \cite{mullick2019generative} and DeepSMOTE \cite{dablain2022deepsmote}. However, these methods have limitations with generating realistic minority class samples when only limited training examples are available. This motivates investigating alternative methods for synthetic image generation, augmentation  and classification (Figure~\ref{fig:pipeline_overview}).

\vspace{1mm}

Quantum Computing (QC) can potentially provide advantages for certain computational tasks through the use of quantum-mechanics principles of superposition and entanglement. Quantum machine learning (QML) provides alternative approaches for representing and learning complex data distributions. Previous work \cite{matondomvula2024breast,azevedo2022quantum,amin2022breast,guha2025resq} has demonstrated various QML algorithms for breast cancer diagnosis. Quanvolutional neural networks have been investigated for breast ultrasound image classification \cite{matondomvula2024breast}, while quantum transfer learning has been applied to mammographic breast cancer detection \cite{azevedo2022quantum}. Hybrid quantum-classical approaches have also addressed breast cancer segmentation \cite{amin2022breast} and histopathology image classification \cite{guha2025resq,goyal2025quantum}. 
Siri \textit{et al.} applied a hybrid deep-learning/quantum approach to an imbalanced breast cancer dataset \cite{siri2025deep}, while Sheoran \textit{et al.} evaluated quantum support vector machine (QSVM), quantum k-nearest neighbors (QKNN), and variational quantum classifier (VQC) methods under class imbalance \cite{sheoran2025robust}. Despite these efforts, the use of quantum generative models to generate additional breast ultrasound samples remains challenging and relatively unexplored. These challenges motivate the use of Quantum Circuit Born Machines (QCBMs) \cite{liu2018differentiable,deluca2026trainability,recioarmengol2025train} that can enable effective generative learning when available training data is limited.

In this paper, we propose a hybrid QML approach using a QCBM to generate realistic synthetic cancer images to augment and balance breast ultrasound training data, as shown in Figure~\ref{fig:pipeline_overview}. In particular, we use a variational autoencoder (VAE) to compress breast ultrasound images into latent vectors and a QCBM with an Instantaneous Quantum Polynomial (IQP) ansatz to learn the distribution of these compressed representations and generate new latent vectors. 
QCBMs use the Born rule to model target data distributions \cite{liu2018differentiable,deluca2026trainability} and can outperform classical generative models in some data-limited settings \cite{hibatallah2024framework}, making them useful for data augmentation.
The generated latent vectors are subsequently decoded into full-resolution synthetic breast ultrasound images that are used for classifier training. 

\begin{figure}[htb]
    \centering
    \subfloat[]{%
        \includegraphics[width=0.51\linewidth]{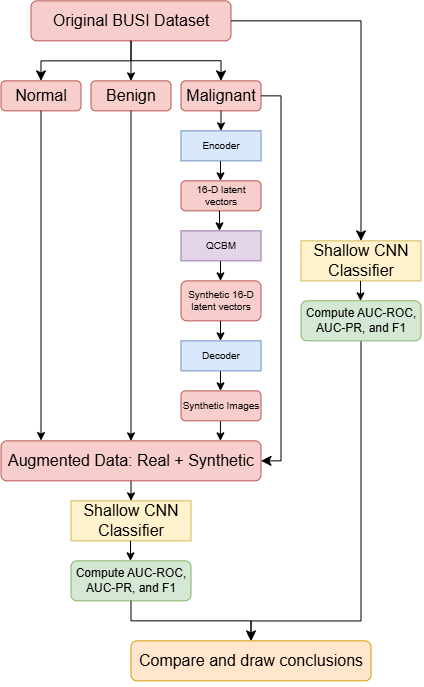}
        \label{fig:pipeline_a}
    }
    \hfill
    \subfloat[]{%
        \includegraphics[width=0.37\linewidth]{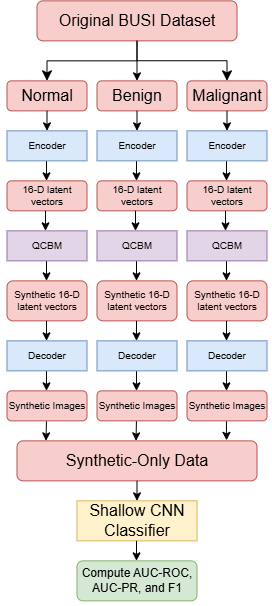}
        \label{fig:pipeline_b}
    }
    \caption{Overview of the proposed pipelines. (a) Augmented method (training the classifier with real and synthetic data ) (b) Synthetic-only method (training the classifier on synthetic data only).
    }
    \label{fig:pipelines}
\end{figure}

We propose two methods, namely an \emph{augmented method} and a \emph{synthetic-only method}, as illustrated in Figure~\ref{fig:pipelines}. For the augmented method, the VAE compresses only malignant training images into latent vectors and trains a QCBM to generate new latent vectors. These vectors are then decoded into synthetic images and combined with the real training data to train the classifier. For the synthetic-only method, separate VAEs and QCBMs are trained to generate synthetic images for each class. The final classifier is trained using only synthetic data. We compare both approaches with a baseline classifier trained using only real images. More specifically, classifier performance is evaluated under three distinct training conditions, namely, real images only, synthetic images only, and a combination of real and synthetic images. Figure~\ref{fig:pipelines} represents pipelines for the two proposed methods.

The main contributions of this work are:
\begin{enumerate}
    \item We propose a hybrid VAE-QCBM framework for generating synthetic breast ultrasound data in a compressed latent space and investigate two training strategies: (i) minority-class augmentation, and (ii) class-specific synthetic-only training.

    \item 
    We evaluate the proposed methods against a real-data baseline and perform an ablation study to isolate the effects of the QCBM from the VAE reconstruction process. 

    \item 	
    We designed a unique 16-qubit quantum circuit and trained it to learn the distribution of breast cancer latent vectors, enabling the generation of new synthetic latent vectors from the learned distribution.

    \item 
    We demonstrate that a classifier trained entirely on QCBM-generated synthetic data achieves a 4\% increase in AUC-ROC while maintaining approximately the same AUC-PR as the real-data baseline.
\end{enumerate}

Our results suggest that QCBM-based hybrid quantum generative methods have strong potential to address class imbalance for breast cancer detection. The remainder of this paper is organized as follows. Section~\ref{sec:methods} provides details on the dataset, algorithm pipeline and our quantum circuit design. Section~\ref{sec:results} provides a discussion of our classification results. Section~\ref{sec:hardware} describes briefly a quantum hardware implementation of our algorithm and Section~\ref{sec:conclusions} summarizes our conclusion.

\section{Dataset and Algorithm Pipeline}
\label{sec:methods}
\vspace{-2mm}
\subsection{Dataset}
\label{subsec:dataset}
\vspace{-2mm}
In our study, we use the BUSI dataset \cite{khaledyan2023enhancing}. The BUSI dataset is challenging due to its small size and due to the speckle nature of ultrasound images \cite{khaledyan2023enhancing}. It consists of 780 256×256 grayscale images out of which 210 are malignant, 437 are benign, and 133 are normal examples. We formulated binary classification as malignant versus non-malignant and randomly partitioned each class into 64\% training, 16\% validation, and 20\% test sets. The split was held fixed across all experiments.

\subsection{VAE Latent Representation}
\label{subsec:vae}
\begin{figure}[htb]
    \includegraphics[width=0.48\textwidth,height=3.4cm]{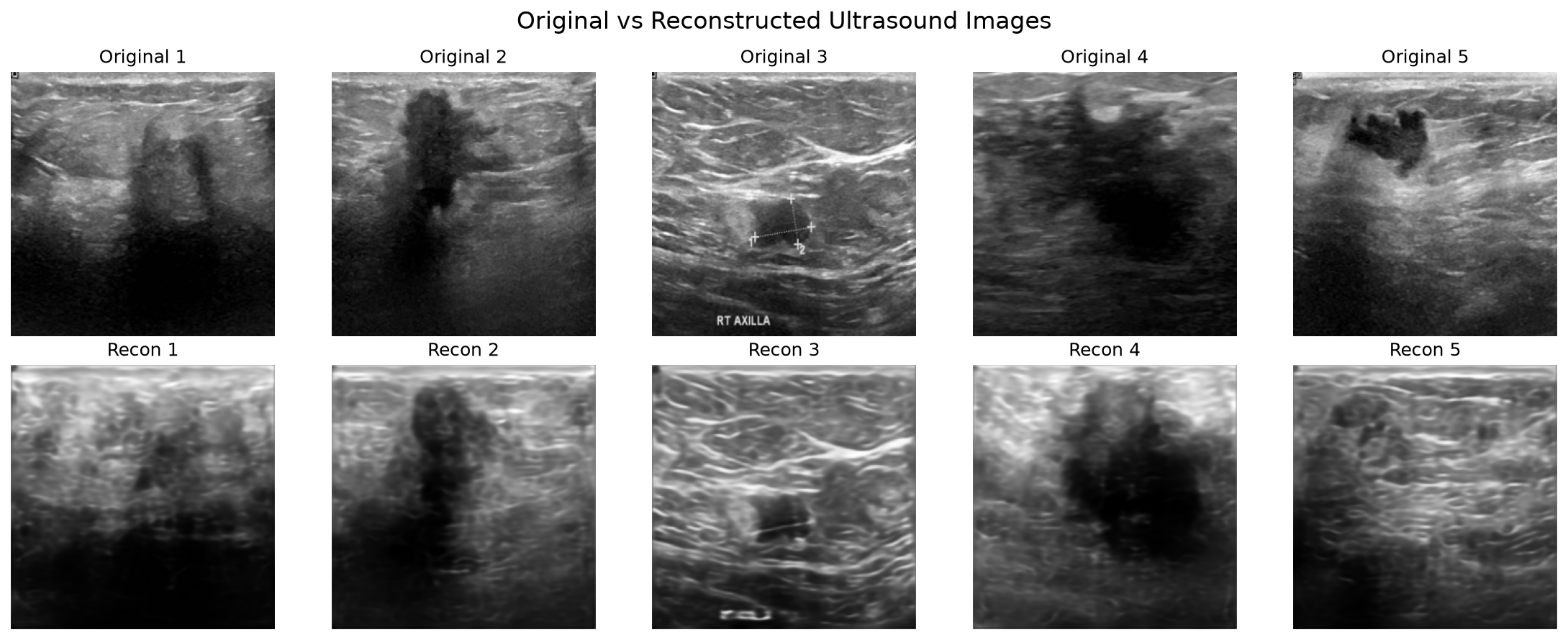}
    \caption{Image comparison; original shown on top versus VAE reconstructed images for the malignant class shown at the bottom.}
    \label{fig:recon}
\end{figure}

\vspace{-2mm}
In our algorithm pipeline, we first compress images into 16-dimensional latent vectors using a VAE. The encoder contains four stride-two convolutional stages with residual blocks and spatial attention, followed by linear heads for the latent mean and log-variance. The decoder expands the latent vector and applies four bilinear upsampling stages with convolutional and residual blocks, spatial attention in the final two stages, and a \(1\times1\) convolution with sigmoid output. The VAE contains approximately 70M parameters. It was trained for 200 epochs using Adam with learning rate \(10^{-4}\), KL weight \(10^{-4}\), and an MSE, \(1-\mathrm{SSIM}\), and KL objective. The model was selected using validation SSIM and visual inspection (Figure~\ref{fig:recon}). We extracted the posterior means, normalized each latent dimension to \([0,1]\), and thresholded at 0.5 to obtain 16-bit strings.

\vspace{-4mm}
\subsection{Quantum Circuit Born Machine}
\label{subsec:qcbm}
\begin{figure}[htb]
\includegraphics[width=\columnwidth]{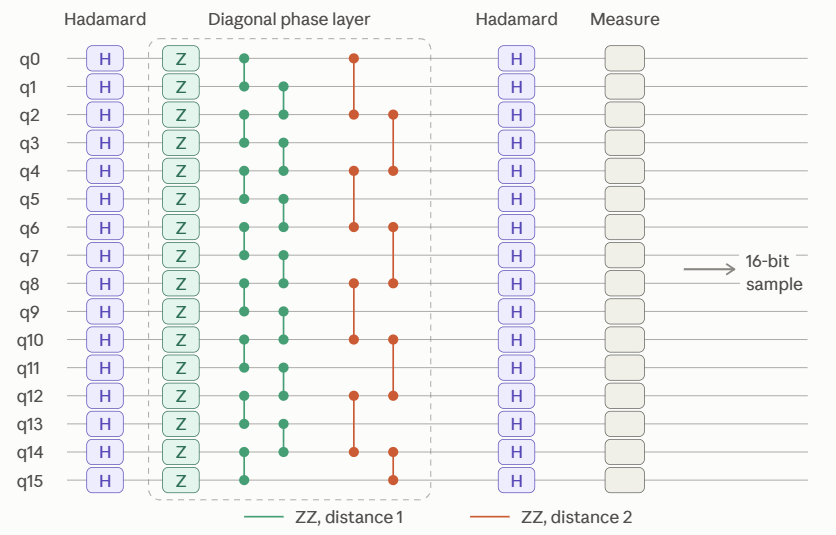}
\caption{Design of a 16-qubit IQP circuit used in QCBM.}
\label{fig:qcirc}
\end{figure}

\vspace{-2mm}
Using the VAE, we encoded our training images into 16-dimensional latents. These serve as the input to train the QCBM using angle encoding. Our QCBM uses a unique 16-qubit instantaneous quantum polynomial (IQP) circuit and was trained for 300 epochs by minimizing the maximum mean discrepancy (MMD) with a Gaussian kernel on Hamming distance, with bandwidth $\sigma$ \cite{liu2018differentiable}. We designed a unique 16-qubit IQP circuit, because it allows for efficient classical simulation via the fast Walsh-Hadamard transform, and hence efficient training. 
While some IQP distributions are hard to sample with classical computation \cite{bremner2016average}, their expectation values can be computed classically \cite{vandenNest2009simulating}, enabling training with MMD loss. Inference can then be performed on quantum hardware \cite{recioarmengol2025train}. 
As shown in Figure~\ref{fig:qcirc}, the circuit consists of a layer of diagonal phase gates sandwiched between layers of Hadamard gates applied to all qubits. The diagonal layer comprises 16 trainable single-qubit \(Z\) gates and 32 trainable two-qubit \(ZZ\) gates, with each qubit entangled with its four nearest neighbors. A fully connected IQP circuit, in which each qubit was entangled with every other qubit via \(ZZ\) gates, exhibited barren-plateau behavior, with the (MMD)\(^2\) loss remaining at approximately \(0.55\). Thus we restricted the entanglement of each qubit to its four nearest neighbors, consistent with \cite{deluca2026trainability}. This modification reduced the (MMD)\(^2\) loss to approximately \(0.01\). We compared the resulting QCBM-generated distribution over bitstrings with the corresponding training distribution. We selected the value of \(\sigma\) that best balanced substantial overlap with the training distribution against a low rate of exact bitstring reproduction. For the remainder of the experiment, we used the QCBM trained with \(\sigma = 1.0\) for the malignant class. Generated bits were mapped to bin centers 0.25 and 0.75, denormalized using the training-set minima and maxima, and decoded by the frozen VAE. 

\vspace{-4mm}
\subsection{Classifier and Evaluation}
\label{subsec:clf}
\vspace{-2mm}
The final classifier was a 25{,}633-parameter CNN with three blocks of convolution, batch normalization, ReLU activation, and max pooling ($1\rightarrow16\rightarrow32\rightarrow64$ channels). This is followed by global average pooling and a dropout-regularized fully connected head. This architecture and its hyperparameters, training for 30 epochs, batch size 16, learning rate $10^{-4}$, and dropout 0.5, are held fixed across every experimental condition reported below.


We report results for four arms of the experiment including baseline, class-weighted, augmented, and synthetic-only. The baseline used real images only. The class-weighted arm \cite{johnson2019survey} also used real images only, with a positive-class weight of approximately 2.7. For the augmented arm, 229 malignant images were generated to obtain a 50/50 malignant versus non-malignant training set. 
For the synthetic-only arm, we generated 500 malignant, 250 benign, and 250 normal images and trained the same classifier on these 1000 synthetic images. All four arms were trained across ten seeds, varying weight initialization, batch ordering, and dropout masks. All arms were evaluated only on real images. Model selection used validation AUC-ROC, and evaluation used the same 157 real test images, including 42 malignant cases. We report AUC-ROC, AUC-PR, F1, and accuracy, treating AUC-ROC and AUC-PR as primary. Paired comparisons used the Wilcoxon signed-rank test; \(p\)-values were not corrected for multiple comparisons.

\section{Results and Analysis}
\label{sec:results}
\vspace{-1mm}
\subsection{Classification Performance}
\label{sec:performance}
\vspace{-2mm}
Table~\ref{tab:results} reports test-set performance. The synthetic-only
arm achieved the highest AUC-ROC, \(0.802 \pm 0.030\) (approximately 80.2\%), compared with
\(0.760 \pm 0.008\) (approximately 76.0\%) for the baseline (mean difference \(+0.042\),
\(p=0.002\)). Its AUC-PR, \(0.645 \pm 0.049\) (approximately 64.5\%), remained comparable to
the baseline value of \(0.643 \pm 0.024\) (approximately 64.3\%). This suggests that the model trained with synthetic only images ranks real breast ultrasound images more accurately, but there was no improvement in retrieval of malignant cases, which are clinically the most relevant.

\vspace{-2mm}

\begin{table}[htb]
\centering
\caption{Test-set performance (mean $\pm$ standard deviation across ten
seeds). All arms share the same classifier, hyperparameters, and real test set.}
\vspace{2mm}
\setlength{\tabcolsep}{3pt}
\renewcommand{\arraystretch}{1.1}
\begin{tabular}{@{}lcccc@{}}
\hline
Arm & AUC-ROC & AUC-PR & F1 & Accuracy \\
\hline
Baseline        & .760$\pm$.008 & .643$\pm$.024 & .070$\pm$.080 & .743$\pm$.012 \\
Class-weighted  & .751$\pm$.006 & .632$\pm$.026 & .577$\pm$.014 & .752$\pm$.026 \\
Augmented       & .752$\pm$.010 & .604$\pm$.034 & .221$\pm$.047 & .754$\pm$.008 \\
Synthetic-only  & .802$\pm$.030 & .645$\pm$.049 & .484$\pm$.147 & .694$\pm$.155 \\
\hline
\label{tab:results}
\end{tabular}
\end{table}

\vspace{-6mm}
\subsection{Analysis of the Augmented Method}
\label{sec:results_aug}
\vspace{-2mm}
The augmented arm did not improve AUC-ROC and reduced AUC-PR relative to the baseline and the class-weighted arm. Under the baseline classifier, real malignant images produced predicted probabilities with mean \(0.365\), standard deviation \(0.1115\), and range \([0.1125,0.5095]\), whereas synthetic malignant images produced mean \(0.5143\), standard deviation \(0.0070\), and range \([0.4831,0.5279]\). The synthetic malignant images were not assigned lower malignancy probabilities, but their predicted probabilities had approximately 16 times less variation as compared to real malignant images. From the perspective of a classifier trained on real data, the 229 synthetic images therefore produced nearly identical responses and represented little variation in classifier-relevant features. Consequently, increasing the number of malignant training images did not necessarily provide an equivalent increase in independent malignant-class information. 

The narrow response distribution could originate at several points in
the generation pipeline. The QCBM could produce insufficiently diverse samples, binarization could remove variation from the sampled latents, or the VAE representation and decoder could map different latents to images containing little classifier-relevant variation. To investigate the bottleneck in the augmented arm, we used the baseline classifier as a fixed probe and compared its predicted malignancy probabilities for real and synthetic malignant images (Table~\ref{tab:bottleneck}). We tested two latent perturbations: temperature \(2.0\) with Gaussian noise standard deviation \(0.05\), and temperature \(3.0\) with noise standard deviation \(0.1\). Neither setting broadened the classifier-score distribution, and increasing the strength of the perturbation produced little additional change. We also replaced the QCBM with a GMM fitted to the same normalized malignant latents and frozen decoder. We sampled 229 latents and evaluated a binarized condition and a continuous condition that bypassed binarization. Changing the sampler, perturbing the latents, and bypassing binarization did not increase classifier-relevant variation after decoding. 
These modifications operated upstream of the decoder and therefore could not restore latent variation downstream. These results indicate that the primary loss of classifier-relevant variation occurs downstream of QCBM sampling and binarization. Since the VAE latent representation and frozen decoder are shared across these conditions, they are the most likely remaining source of the bottleneck. 


\begin{table}[htb]
\centering
\caption{Bottleneck analysis based on predicted malignancy probability distributions.}
\vspace{2mm}
\label{tab:bottleneck}
\begin{tabular}{lc}
\hline
\textbf{Condition} & \textbf{Std. Dev.} \\
\hline
Real malignant images              & 0.1115 \\
QCBM                               & 0.0070 \\
QCBM ($T=2.0$, noise $=0.05$)      & 0.0053 \\
QCBM ($T=3.0$, noise $=0.10$)      & 0.0056 \\
GMM (binarized)                    & 0.0065 \\
GMM (continuous)                   & 0.0052 \\
\hline
\end{tabular}
\end{table}

\vspace{-4mm}
\subsection{Analysis of the Synthetic Method}
\label{sec:results_aug}
\vspace{-2mm}
In the augmented arm, synthetic images were generated only for the malignant class, creating an association between synthetic origin and the positive label. This motivated training entirely on synthetic images generated separately for the malignant, benign, and normal classes. The synthetic-only arm removes the real-versus-synthetic distinction. The final classification task remains binary, but the generation stage now models all three source classes rather than only the malignant distribution. Since separate VAEs may introduce class-specific patterns, we tested whether VAE identity was recoverable from the reconstructions using a cross-reconstruction probe. The same 283 real validation and test images were reconstructed using all three class-specific VAEs, producing 849 cross-reconstructions. An equal-capacity classifier was then trained to predict the generating VAE. Splits were made by source image, with 85 source images, or 255 reconstructions, held out per run. Since the same source images were used for all three VAEs, accuracy above the \(33.3\%\) chance level indicates a recoverable VAE-specific signature. The probe achieved a median accuracy of \(96.1\%\) across ten runs, ranging from \(0.667\) to \(1.000\). The malignant VAE was identified correctly in 668 of 849 cases, compared with 796 of 849 for benign and 849 of 849 for normal. These results show that class-specific VAE signatures are present and could provide a shortcut for the augmented and synthetic-only classifiers. However, the probe was explicitly trained to recover VAE identity, so it does not establish that the malignancy classifiers used this information. Evaluation was performed only on real images, which contain no VAE fingerprint, so such a shortcut cannot transfer directly to the test set.

The resulting AUC-ROC increase in the synthetic arm evaluated on the real test set (Table~\ref{tab:results}) provides evidence that images from the hybrid quantum pipelines can support learning that transfers to real images. These results suggest that modeling malignant, benign, and normal distributions separately within the hybrid quantum-classical generation pipeline can provide useful information for improving ranking for malignant versus non-malignant cases.

\vspace{-2mm}
\section{Quantum Hardware Evaluation}
\label{sec:hardware}
\begin{figure}[htb]
    \centering
    \includegraphics[width=0.75\columnwidth]{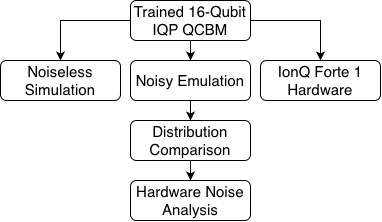}
    \caption{Quantum hardware evaluation pipeline. The trained IQP QCBM is sampled using noiseless simulation, noisy emulation, and the IonQ Forte 1 processor. The resulting distributions are compared to characterize hardware-induced deviations.}
    \label{fig:hardware_implementation}
\end{figure}

\vspace{-2mm}
The QCBM used in this work is designed to support execution on quantum hardware in addition to classical simulation. As shown in Figure~\ref{fig:hardware_implementation}, the trained 16-qubit IQP QCBM will be evaluated using noisy emulation and physical quantum hardware \cite{goyal2026configurable}. The circuit will be executed for sampling on the IonQ Forte 1 trapped-ion processor through Amazon Braket, and the resulting sample distribution will be compared with the noiseless simulated QCBM distribution used in this study. Hardware execution introduces finite-shot noise, gate and readout errors, decoherence, and transpilation overhead, which can alter the sampled bitstring distribution and the latent vectors passed to the VAE decoder. Hardware-specific noise characterization, the noise effect on the generated distribution, and applicable error-mitigation strategies will be documented in the final paper. Detailed comparison of hardware, simulated QCBM, and malignant-data distributions will also be presented in the final paper.

\section{Conclusion}
\label{sec:conclusions}
\vspace{-2mm}
In this paper, we developed a QML algorithm and employed generative methods to address the database imbalance problem in breast cancer detection and classification studies. In addition, we designed a 16-qubit quantum circuit and evaluated two augmentation pipelines for an imbalanced breast ultrasound dataset. Augmenting the real training set with QCBM-generated malignant images produced no improvement over a baseline trained on real images alone. The baseline classifier showed that the synthetic images received near-identical malignancy probabilities. In addition, ablation tests identified the VAE representation and decoder, rather than QCBM sampling, as the likely source of this limited variation. Since synthetic images were generated only for the malignant class, synthetic origin was also associated with the positive label. This motivated the synthetic-only arm, in which separate VAEs and QCBMs generated malignant, benign, and normal images and the classifier was trained entirely on synthetic data. This arm achieved a 4\% higher AUC-ROC score over the real-data baseline while maintaining comparable AUC-PR. The cross-reconstruction probe showed that VAE-specific signatures were present in the generated images, although this does not establish that the malignancy classifier relied on these signatures. Evaluation was performed only on real test images, therefore the observed AUC-ROC improvement reflects performance that transferred from synthetic training data to real breast ultrasound images. We note that we also developed the entire algorithm for implementation on quantum hardware, namely, on the IonQ Forte 1 trapped-ion processor. This will allow assessing the performance of our algorithm with quantum noise along with comparisons against noiseless simulations. The hardware implementation along with associated quantum hardware results will be described in more detail in the final paper and in the conference presentation.


\vfill

\bibliographystyle{IEEEbib}
\bibliography{refs}

@article{venkatesh2015going,
  author  = {S. S. Venkatesh and B. J. Levenback and L. R. Sultan and G. Bouzghar and C. M. Sehgal},
  title   = {Going beyond a first reader: A machine learning methodology for optimizing cost and performance in breast ultrasound diagnosis},
  journal = {Ultrasound in Medicine \& Biology},
  volume  = {41},
  number  = {12},
  pages   = {3148--3162},
  year    = {2015},
}

@article{mahmood2020brief,
  author  = {T. Mahmood and others},
  title   = {A brief survey on breast cancer diagnostic with deep learning schemes using multi-image modalities},
  journal = {IEEE Access},
  volume  = {8},
  pages   = {165779--165809},
  year    = {2020}
}

@article{mambou2018breast,
  author  = {S. J. Mambou and P. Maresova and O. Krejcar and A. Selamat and K. Kuca},
  title   = {Breast cancer detection using infrared thermal imaging and a deep learning model},
  journal = {Sensors},
  volume  = {18},
  number  = {9},
  pages   = {2799},
  year    = {2018},
}

@inproceedings{aboudessouki2025enhanced,
  author    = {A. Aboudessouki and others},
  title     = {Enhanced breast cancer molecular biomarker classification: A novel two-stage machine learning pipeline for accurate histological analysis of whole slide images},
  booktitle = {IEEE ICASSP},
  pages     = {1--5},
  year      = {2025},
  
}

@article{lopez2023deep,
  author  = {M. L{\'o}pez-P{\'e}rez and P. Morales-{\'A}lvarez and L. A. D. Cooper and R. Molina and A. K. Katsaggelos},
  title   = {Deep Gaussian processes for classification with multiple noisy annotators: Application to breast cancer tissue classification},
  journal = {IEEE Access},
  volume  = {11},
  pages   = {6922--6934},
  year    = {2023}
}

@article{loizidou2025subtraction,
  author  = {K. Loizidou and G. Skouroumouni and G. Savvidou and A. Constantinidou and E. Vlachou and A. Yiallourou and C. Pitris and C. Nikolaou},
  title   = {Subtraction of temporally sequential digital mammograms: Enhancing the detection and classification of malignant masses in breast imaging},
  journal = {IEEE JIIM},
  note = {38, (2025): 4243-4254},
}

@inproceedings{liasis2012combination,
  author    = {G. Liasis and C. Pattichis and S. Petroudi},
  title     = {Combination of different texture features for mammographic breast density classification},
  booktitle = {IEEE BIBE},
  pages     = {732--737},
  year      = {2012},
  publisher = {IEEE}
}

@inproceedings{sun2004two,
  author    = {Y. Sun and C. F. Babbs and E. J. Delp III},
  title     = {A two-stage classifier system for normal mammogram identification},
  booktitle = {Computational Imaging II},
  volume    = {5299},
  pages     = {112--122},
  publisher = {SPIE},
  year      = {2004}
}

@article{liu2025famf,
  author  = {Y. Liu and others},
  title   = {{FAMF-Net}: Feature alignment mutual attention fusion with region awareness for breast cancer diagnosis via imbalanced data},
  journal = {IEEE Trans. on Medical Imaging},
  volume  = {44},
  number  = {3},
  pages   = {1153--1167},
  month   = mar,
  year    = {2025},
}

@article{wu2023cross,
  author  = {H. Wu and X. Huang and X. Guo and Z. Wen and J. Qin},
  title   = {Cross-image dependency modeling for breast ultrasound segmentation},
  journal = {IEEE Trans. on Medical Imaging},
  volume  = {42},
  number  = {6},
  pages   = {1619--1631},
  month   = jun,
  year    = {2023}
}

@article{jiang2026semi,
  author  = {S. Jiang and H. Wu and Y. Zhou and J. Chen and J. Qin},
  title   = {Semi-supervised breast lesion segmentation using confidence-ranked features and bi-level prototypes},
  journal = {IEEE Trans. Neural Netw. Learn. Syst.},
  volume  = {37},
  number  = {3},
  pages   = {1290--1301},
  month   = mar,
  year    = {2026}
}

@article{muralidhar2008computer,
  author  = {G. S. Muralidhar and T. M. Haygood and T. W. Stephens and G. J. Whitman and A. C. Bovik and M. K. Markey},
  title   = {Computer-aided detection of breast cancer---Have all bases been covered?},
  journal = {Breast Cancer: Basic and Clinical Research},
  volume  = {2},
  pages   = {5--9},
  year    = {2008},
}

@article{li2019discriminative,
  author  = {X. Li and M. Radulovic and K. Kanjer and K. N. Plataniotis},
  title   = {Discriminative pattern mining for breast cancer histopathology image classification via fully convolutional autoencoder},
  journal = {IEEE Access},
  volume  = {7},
  pages   = {36433--36445},
  year    = {2019},
}

@incollection{pattichis2002modular,
  author    = {Pattichis, C. S. and Schnorrenberg, F. and Schizas, C. N. and Pattichis, M. S. and Kyriacou, K.},
  title     = {A Modular Neural Network System for the Analysis of Nuclei in Histopathological Sections},
  booktitle = {Computational Intelligence Processing in Medical Diagnosis},
  pages     = {291--322},
  publisher = {Physica-Verlag HD},
  address   = {Heidelberg, Germany},
  year      = {2002}
}

@article{matondomvula2024breast,
  author  = {N. Matondo-Mvula and K. Elleithy},
  title   = {Breast cancer detection with quanvolutional neural networks},
  journal = {Entropy},
  volume  = {26},
  number  = {8},
  pages   = {630},
  year    = {2024}
}

@article{azevedo2022quantum,
  author  = {V. Azevedo and C. Silva and I. Dutra},
  title   = {Quantum transfer learning for breast cancer detection},
  journal = {Quantum Machine Intelligence},
  volume  = {4},
  number  = {1},
  pages   = {5},
  year    = {2022},
}

@article{amin2022breast,
  author  = {J. Amin and others},
  title   = {Breast microscopic cancer segmentation and classification using unique 4-qubit-quantum model},
  journal = {Microsc. Res. Tech.},
  volume  = {85},
  number  = {5},
  pages   = {1926},
  year    = {2022},
}

@article{guha2025resq,
  author  = {D. Guha and S. Kuiry and S. Mitra and S. Bhattacharyya and N. Das},
  title   = {{ResQ}: A hybrid classical-quantum model for efficient breast cancer image classification},
  journal = {Applied Soft Computing},
  volume  = {183},
  pages   = {113631},
  month   = nov,
  year    = {2025},
}

@inproceedings{mullick2019generative,
  author    = {S. S. Mullick and S. Datta and S. Das},
  title     = {Generative adversarial minority oversampling},
  booktitle = {IEEE ICCV},
  pages     = {1695--1704},
  year      = {2019}
}

@article{dablain2022deepsmote,
  author  = {D. Dablain and B. Krawczyk and N. V. Chawla},
  title   = {{DeepSMOTE}: Fusing deep learning and {SMOTE} for imbalanced data},
  journal = {IEEE Trans. Neural Netw. Learn. Syst.},
  volume  = {34},
  number  = {9},
  pages   = {6390},
  year    = {2022}
}

@inproceedings{siri2025deep,
  author    = {P. Siri and G. Anitha and M. R. N. V. Durga},
  title     = {A deep learning driven quantum hybrid approach for early breast cancer detection},
  booktitle = {IEEE AISP},
  pages     = {1--5},
  year      = {2025},
}

@article{sheoran2025robust,
  author  = {S. K. Sheoran and V. Yadav and R. K. Sheoran},
  title   = {Robust evaluation of classical and quantum machine learning under noise, imbalance, feature reduction and explainability},
  journal = {Scientific Reports},
  volume  = {15},
  pages   = {45714},
  month   = dec,
  year    = {2025},
}

@article{deluca2026trainability,
  author  = {G. De Luca},
  title   = {Trainability of {IQP} quantum circuit born machines under Gaussian initialization},
  journal = {arXiv preprint arXiv:2606.10179},
  year    = {2026}
}

@article{hibatallah2024framework,
  author  = {M. Hibat-Allah and M. Mauri and J. Carrasquilla and A. Perdomo-Ortiz},
  title   = {A framework for demonstrating practical quantum advantage: Comparing quantum against classical generative models},
  journal = {Communications Physics},
  volume  = {7},
  number  = {1},
  pages   = {68},
  year    = {2024}
}

@article{khaledyan2023enhancing,
  author  = {D. Khaledyan and T. J. Marini and T. M. Baran and A. O'Connell and K. Parker},
  title   = {Enhancing breast ultrasound segmentation through fine-tuning and optimization techniques: Sharp attention {U-Net}},
  journal = {PLOS ONE},
  volume  = {18},
  number  = {12},
  pages   = {e0289195},
  year    = {2023}
}

@article{liu2018differentiable,
  author  = {J.-G. Liu and L. Wang},
  title   = {Differentiable learning of quantum circuit born machines},
  journal = {Physical Review A},
  volume  = {98},
  number  = {6},
  pages   = {062324},
  year    = {2018}
}

@article{recioarmengol2025train,
  author  = {E. Recio-Armengol and S. Ahmed and J. Bowles},
  title   = {Train on classical, deploy on quantum: Scaling generative quantum machine learning to a thousand qubits},
  journal = {arXiv preprint arXiv:2503.02934},
  year    = {2025}
}

@article{johnson2019survey,
  author  = {J. Johnson and T. Khoshgoftaar},
  title   = {Survey on deep learning with class imbalance},
  journal = {J. Big Data},
  volume  = {6},
  number  = {1},
  pages   = {27},
  year    = {2019}
}

@inproceedings{goyal2025quantum,
  author    = {N. Goyal and G. Uehara and A. Spanias},
  title     = {Quantum-Enhanced Cancer Detection for Histopathologic Images},
  booktitle = {Proc. IEEE Int. Conf. Image Process. (ICIP)},
  year      = {2025},
}

@inproceedings{goyal2026configurable,
  author    = {N. Goyal and G. Uehara and A. Spanias},
  title     = {Configurable Algorithms for Histopathologic Cancer Detection on Quantum Hardware},
  booktitle = {Proc. Asilomar Conf. Signals, Syst., Comput.},
  year      = {2026}
}

@article{bremner2016average,
  author  = {Bremner, Michael J. and Montanaro, Ashley and Shepherd, Dan J.},
  title   = {Average-case complexity versus approximate simulation of commuting quantum computations},
  journal = {Physical Review Letters},
  volume  = {117},
  number  = {8},
  pages   = {080501},
  year    = {2016},
}

@misc{vandenNest2009simulating,
  author        = {Van den Nest, M.},
  title         = {Simulating quantum computers with probabilistic methods},
  year          = {2009},
  eprint        = {0911.1624},
  archivePrefix = {arXiv},
  primaryClass  = {quant-ph},
}

\end{document}